\documentclass[12pt,aps,prb,preprint]{revtex4}
\usepackage{graphicx}
\usepackage{psfrag}
\usepackage{amsmath, amssymb}
\usepackage{amsthm}

\usepackage{mathpazo} 
\usepackage[scaled]{helvet} 
\usepackage{courier} 
\normalfont

\newcommand{\be}{\begin{equation}}
\newcommand{\ee}{\end{equation}}
\newcommand{\bes}{\begin{equation*}}
\newcommand{\ees}{\end{equation*}}

\theoremstyle{definition}

\theoremstyle{remark}

\begin{document}

\title{The mystery of square root of minus one in quantum mechanics,\\and its demystification}
\author{C.P. Kwong}
\affiliation{The Chinese University of Hong Kong, Department of Mechanical and Automation Engineering, Shatin, N.T., Hong Kong}
\email{cpkwong@mae.cuhk.edu.hk}
\date{December 20, 2009}      

\begin{abstract}
To most physicists, quantum mechanics must embrace the imaginary number $i\!=\!\sqrt{-1}$ is at least a common belief if not a mystery.    We use the famous example $\mathbf{pq}-\mathbf{qp}=\frac{h}{2\pi i}\mathbf{I}$ to demonstrate the possible elimination of $i$ when constructing this noncommutative relationship.  We then discuss the role of $i$ in the formulation of Schr{\"o}dinger's wave equation.  Common to the original development of these two quantum theories was the use of complex exponential to represent the fundamental variables (i.e., $\mathbf{p}$, $\mathbf{q}$, and the wave function).  Understanding this complex function from the right perspective, as we suggest in this essay, removes the mysteries surrounding the complex nature of quantum mechanics. 
\end{abstract}    

\maketitle

\section{Introduction} The imaginary number~~$i\!=\!\sqrt{1}$~~emerged in all orthodox models of quantum mechanics: first the matrix mechanics of Born and Jordan, then the Poisson bracket of Dirac and the Schr{\"o}dinger equation, and later the Feynman's path integral.  It appears that $i$ is \textbf{\emph{indispensable}} in the formulation of quantum mechanics.  This last assertion was advocated seriously or otherwise.  For the former, we refer to a seminal presentation by C.N. Yang in celebrating the centenary of the birth of Schr{\"o}dinger. \citep{yang87} For the latter, we quote two examples one drawn from a widely-circulated article by E.P. Wigner published in 1960 and the other a recent lecture by F. Dyson: 

\begin{quote}
Surely to the unpreoccupied mind, complex numbers are far from natural or simple and they cannot be suggested by physical observations. Furthermore, the use of complex numbers is in this case not a calculational trick of applied mathematics but comes close to being a necessity in the formulation of the laws of quantum mechanics.\cite{wigner60}
\end{quote}
\begin{quote}
$\dotsc$But then came the surprise. Schr{\"o}dinger put the square root of minus one into the equation, and suddenly it made sense.  Suddenly it became a wave equation instead of a heat conduction equation. $\dotsc$And that square root of minus one means that nature works with complex numbers and not with real numbers.\cite{dyson09}
\end{quote}

We show in this short article that the introduction of the imaginary number $i$ to Born-Jordan's formulation of quantum mechanics is \textbf{\emph{inessential}} though it enables a compact representation of the noncommutative relationship between momentums and positions.  Our arguments are based simply on the fundamental understanding of the three \textbf{\emph{equivalent}} forms of Fourier series.  We will review the development of these forms and then derive a new version of the above noncommutative relationship which contains only real elements.  We then move on to examine the role of $i$ in the time-dependent Schr{\"o}dinger equation following the line of thought of Schr{\"o}dinger.  We argue that the equation is necessarily complex only because we demand a complex solution.  Furthermore, a complex wave function in the form of a complex exponential encodes no more information than a real sinusoidal wave function.  Thus the introduction of the imaginary number $i$ to the Schr{\"o}dinger equation is again for the sake of compact representation.

Reflection on removing the imaginary element from physics is not new though scanty.\cite{gull&etal93,hestenes86} However, the studies have drawn little interest from the majority of physicists, not to mention changing their perspective.  One possible reason is that new mathematics, notably Clifford algebra, is required to understand the underling reasoning of these studies.  On the other hand, the way we approach the problem, albeit its simplicity, suffices to clear the mist created by $\sqrt{-1}$ in quantum mechanics.  Another explanation of the apathy, which we wish it to be wrong, is that physicists ``quickly become impatient with any discussion of elementary concepts,\cite{hestenes86}''  Nevertheless, the persistence of misconception about the role of $i$ in quantum mechanics is undesirable from at least the pedagogical angle.

\section{The three forms of Fourier series}
Given a real periodic function $f(t)$ of period $T_0$, the \textbf{trigonometric Fourier series} expansion of $f(t)$ is
\be\label{tFs}
f(t)=a_0 +\sum_{k=1}^{\infty}\left[a_k \cos(k\omega_0 t)+ b_k
\sin(k\omega_0 t)\right],
\ee                
where $\omega_0=2\pi\nu_0=\frac{2\pi}{T_0}$ is the fundamental frequency and $a_0, a_k$, and $b_k$ are Fourier coefficients determined by, respectively,
\be
a_0=\frac{1}{T_0}\int_{0}^{T_0}f(t)\,dt,
\ee
\be
a_k=\frac{2}{T_0}\int_{0}^{T_0}f(t)\cos(k\omega_0 t)\,dt,
\ee
and
\be
b_k=\frac{2}{T_0}\int_{0}^{T_0}f(t)\sin(k\omega_0 t)\,dt.
\ee

If the periodic $f(t)$ is an even function, i.e.,
\be
f(-t)=f(t),
\ee
$b_k$ are zero for all $k=1,2,\dotsc$.  That means the Fourier series contains only the constant term $a_0$ and the cosine functions.  This can be easily proved mathematically or appreciated intuitively (i.e., it suffices to sum the cosine functions $\cos{(k\omega_0 t)}$, which are even, to give an even function).  On the other hand, for an odd periodic $f(t)$, i.e.,
\be
f(-t)=-f(t),
\ee
$a_k=0$ for all $k=0,1,2,\dotsc$.  In this case, the Fourier series is a sine series.  A general $f(t)$ is neither even nor odd and hence both cosine and sine functions are required.

The second form of Fourier series gives rise the concept of ``phase''.  
We illustrate the idea by considering a simple example of trigonometry:
\be\sin(\omega_0
t)+2\cos(\omega_0 t)=\sqrt{5}\cos(\omega_0 t-26.6^{\circ}).\ee
Thus, the sum of a cosine function and a sine function of the same
frequency can be expressed as a cosine function with the same
frequency but with an added ``phase shift''.  In general, \be a_k
\cos(k\omega_0 t)+b_k\sin(k\omega_0 t)=c_k \cos(k\omega_0
t+\theta_k)\ee where
\be\label{c_cFs}
c_k=\sqrt{{a_k}^{2}+{b_k}^{2}}
\ee
is the amplitude and
\be\label{t_cFs}
\theta_k=\tan^{-1}\frac{-b_k}{a_k}
\ee
is the phase.  Hence the trigonometric Fourier series~\eqref{tFs} can be rewritten as
\be\label{cFs}
f(t)=c_0+\sum_{k=1}^{\infty}c_k \cos(k\omega_0 t+\theta_k)
\ee
with $c_0=a_0$.  We call Equation~\eqref{cFs} the \textbf{Fourier cosine series} of
$f(t)$.

The Euler's equation \be\label{Eeqn} e^{\pm ix}=\cos(x)\pm i \sin(x)\ee  leads us to the third form of Fourier series.  Since, by virtue of~\eqref{Eeqn}, \be\cos(x)=\frac{1}{2}(e^{ix}+e^{-ix})\ee and
\be\sin(x)=\frac{1}{2i}(e^{ix}-e^{-ix}),\ee
we can then rewrite the trigonometric Fourier series~\eqref{tFs} as 
\be\begin{split} f(t) &=a_0+\sum_{k=1}^{\infty}\left(a_k
\frac{e^{ik\omega_0 t}+e^{-ik\omega_0
       t}}{2}+b_k \frac{e^{ik\omega_0 t}-e^{-ik\omega_0 t}}{2i}\right)\\
     &=a_0+\sum_{k=1}^{\infty}\left[\left(\frac{a_k-ib_k}{2}\right) e^{ik\omega_0
       t}+\left(\frac{a_k+ib_k}{2}\right)e^{-ik\omega_0 t}\right].
\end{split}\ee
Let
$\bar c_k=\dfrac{a_k-ib_k}{2}$, $\bar c_{-k}=\dfrac{a_k+ib_k}{2}$, and $\bar c_0=a_0$, we obtain
\be\label{ceFs1}
f(t)=\bar c_0+\sum_{k=1}^{\infty}(\bar c_k e^{ik\omega_0 t}+\bar c_{-k}e^{-ik\omega_0 t}).
\ee
Obviously $\bar{c}_k$ and $\bar{c}_{-k}$ are in general complex. Letting $k$ to run from 1 to $\infty$ in~\eqref{ceFs1} is equivalent to letting $k$ to run from $-\infty$ to $+\infty$ (including zero) in a more compact equation:
\be\label{ceFs2}
f(t)=\sum_{k=-\infty}^{\infty}\bar c_k e^{ik\omega_0 t}.
\ee
Equation~\eqref{ceFs2} is called the \textbf{complex Fourier series}.  In this series the $k$th complex Fourier coefficient $\bar c_k$ can be expressed in its polar form $\bar c_k=|\bar
c_k|e^{i\theta_k}$.  Then $\bar c_{-k}=\bar c_{k}^*$ (complex conjugate of
$\bar c_k$)
$=|\bar c_k|e^{-i\theta_k}$.  It is easy to show that, for all integers $k$ except zero,
\be\label{c_ceFs}
|\bar c_k|=\frac{1}{2}\sqrt{{a_{k}}^{2}+{b_{k}}^{2}}
\ee
and
\be\label{t_ceFs}
\theta_k=\tan^{-1}\frac{-b_k}{a_k}.
\ee
Compare~\eqref{c_cFs} and~\eqref{t_cFs} with~\eqref{c_ceFs} and~\eqref{t_ceFs} we see immediately that the phase (the magnitude) of the complex Fourier series is identical to (half) that of the Fourier cosine series.  Nevertheless, all these quantities are uniquely determined by the Fourier coefficients ($a_0,a_k$, and $b_k$) of the trigonometric Fourier series~\eqref{tFs}.  It is thus obvious that the cosine Fourier series~\eqref{cFs} and the complex Fourier series~\eqref{ceFs2} are merely variations on the \textbf{\emph{same}} theme of the trigonometric Fourier series.

\section{The real noncommutative relationship}
Inspired by the work of Heisenberg,\cite{heisen25} Born and Jordan~\cite{born&jordan25} derived in their historical paper the noncommutative relationship between momentums $p$ and positions $q$:
\be
\mathbf{pq}-\mathbf{qp}=\frac{h}{2\pi i}\mathbf{I}.
\ee 
Since $\mathbf{I}$ is the identity matrix, $\mathbf{pq}-\mathbf{qp}$ is a diagonal matrix of which every diagonal element equals to $\frac{h}{2\pi i}$.  The $n$th diagonal element, expressed in terms of $p$ and $q$, is given by~\cite{born&jordan25}
\be\label{diag_element}
\sum_{k=1}^{\infty}(p_{n,k}q_{k,n}-q_{n,k}p_{k,n}).
\ee  
Therefore the noncommutative relationship dictated by the $n$th diagonal element is
\be\label{complex_noncommu}
\boxed{\sum_{k=1}^{\infty}(p_{n,k}q_{k,n}-q_{n,k}p_{k,n})=\frac{h}{2\pi i}\,.}
\ee
 
One should not be too surprised to learn that the summation \eqref{diag_element} turns out to be complex-valued (more precisely, purely imaginary).  It is because, in the derivation of Born and Jordan,  \textbf{\emph{complex}} Fourier series were used to expand the \textbf{\emph{real}} $p$ and $q$:
\be
p=\sum_{k=-\infty}^{\infty}p_k e^{ik\omega_0 t},\quad q=\sum_{k=-\infty}^{\infty}q_k e^{ik\omega_0 t}.
\ee 
Moreover, for any four complex numbers $\alpha,\beta,\rho,$ and $\sigma$, $\alpha\beta-\rho\sigma=\frac{K}{i}$ ($K$ is a real constant) whenever $\rho=\beta^*$ and $\sigma=\alpha^*$.  The four complex coefficients $p_{n,k},q_{k,n},q_{n,k}$, and $p_{k,n}$ satisfy exactly these conditions to give~\eqref{complex_noncommu}.
As we have shown in the last section, the complex Fourier series~\eqref{ceFs2} is identical to the trigonometric Fourier series~\eqref{tFs}.  It is therefore completely legitimate to expand $p$ and $q$ each into a trigonometric Fourier series, which is \textbf{\emph{real}}:
\begin{gather}\label{tFs_pk}
\begin{split}
p &=p_0+\sum_{k=1}^{\infty}\left[p_k^c\cos(k\omega_0 t)+p_k^s\sin(k\omega_0 t)\right],\\
q &=q_0+\sum_{k=1}^{\infty}\left[q_k^c\cos(k\omega_0 t)+q_k^s\sin(k\omega_0 t)\right].
\end{split}
\end{gather}
In Eq.~\eqref{tFs_pk} the superscript $c$ (superscript $s$) of $p_k$ or $q_k$ signifies that the coefficient is associated with a cosine (sine) component.  In the following derivation we simply replicate the approach of Heisenberg~\cite{heisen25} and Born and Jordan~\cite{born&jordan25}.  Detailed examination of the historical and technical developments of this approach can be found in Refs.~3 and 4.
 
Given $q$ in the form of Eq.~\eqref{tFs_pk}, we have
\be\label{q_dot}
\dot{q}=\sum_{k=1}^{\infty}\left[q_k^s\,k\omega_0\cos(k\omega_0 t)-q_k^c\,k\omega_0\sin(k\omega_0 t)\right].
\ee
For applying the correspondence principle we substitute the trigonometric Fourier expansions of $\dot{q}$ and $p$ into the Bohr-Sommerfeld's quantum condition    
\be\label{quan_cond}
nh=\int_{0}^{\frac{1}{\nu_0}}p\dot{q}\,dt
\ee
where $n$ is the quantum number, $h$ is the Planck's constant, and $\nu_0=\frac{\omega_0}{2\pi}$.  The integration of $p\dot{q}$ over the fundamental period from $0$ to $\frac{1}{\nu_0}$ is much simplified due to the orthogonal properties of sine and cosine functions.  Indeed, it can be easily shown that only integrands associated with $\sin^2(k\omega_{0}t)$ or $\cos^2(k\omega_{0}t)$ will give finite values of integration; the remaining integrands all render the same integration result of zero.  Specifically, we have, for all $k=1,2,\dotsc$,
\be
\int_{0}^{\frac{1}{\nu_0}}\sin^2(k\omega_{0}t)\,dt=\int_{0}^{\frac{1}{\nu_0}}\cos^2(k\omega_{0}t)\,dt=\frac{1}{2\nu_0}.
\ee
Thus the quantum condition~\eqref{quan_cond} becomes
\be\label{quan_cond_Fs}
nh=\pi\sum_{k=1}^{\infty}k(p_k^{c}q_k^{s}-p_k^{s}q_k^{\verb""c}).
\ee
Differentiate Eq.~\eqref{quan_cond_Fs} with respect to $n$ gives
\be
h=\pi\sum_{k=1}^{\infty}k\frac{\partial}{\partial n}(p_k^{c}q_k^{s}-p_k^{s}q_k^{c}).
\ee
The further application of Born's correspondence rule
\be
k\frac{\partial\nu(n,k)}{\partial n}\,\leftrightarrow\,\nu_{n+k,n}-\nu_{n,n-k}
\ee
gives
\be\label{real_noncommu}
\boxed{\sum_{k=1}^{\infty}\left[(p_{n+k,n}^{c}q_{n+k,n}^{s}-p_{n+k,n}^{s}q_{n+k,n}^{c})-(p_{n,n-k}^{c}q_{n,n-k}^{s}-p_{n,n-k}^{s}q_{n,n-k}^{c})\right]=\frac{h}{\pi}.}
\ee
Equation~\eqref{real_noncommu} is our desired \textbf{\emph{real}} noncommutative relationship between momentums and positions.  This is in contrast with its complex counterpart Eq.~\eqref{complex_noncommu}.  Note that Equation~\eqref{real_noncommu} can be simplified when $p=m\dot{q}$ and either $p$ or $q$ is an odd or even function.  Take for example $q$ is an odd function.  Then $q$ can be expressed as a Fourier series having only sine functions.  Since the derivative of an odd function is an even function, then $p$ can be expressed as a Fourier series having only a constant term and cosine functions.  Consequently, the quantum condition Eq.~\eqref{quan_cond_Fs} becomes
\be
nh=m\pi\sum_{k=1}^{\infty}kp_k^{c}q_k^{s}
\ee
and Eq.~\eqref{real_noncommu} reduces to
\be
\sum_{k=1}^{\infty}(p_{n+k,n}^{c}q_{n+k,n}^{s}-p_{n,n-k}^{c}q_{n,n-k}^{s})=\frac{h}{m\pi}.
\ee

\section{``Schr{\"o}dinger equation is complex'' is no mystery}
At a conference celebrating the centenary of the birth of Schr{\"o}dinger, \citep{yang87} C.N. Yang gave an intelligent guess on why Schr{\"o}dinger resisted the imaginary number $i$ at the outset but finally accepted its appearance in his (time-dependent) wave equation we nowadays encounter:
\be\label{SE} i\hbar\frac{\partial}{\partial t}\psi=\left[-\frac{\hbar^2}{2m}\nabla^{2}+V\right]\psi,\ee where $\hbar=\frac{h}{2\pi}$, $V$ is the potential, and \be\nabla^{2}=\frac{\partial^{2}}{\partial x^2}+\frac{\partial^{2}}{\partial y^2}+\frac{\partial^{2}}{\partial z^2}\ee is the Laplace operator.  However, we argue that Schr{\"o}dinger's acceptance of $i$ was hardly wholehearted.  Furthermore, the belief that his equation~\eqref{SE} must be complex is only a myth.     

In contrast with the Schr{\"o}dinger equation the classical equation for modeling waves, such as transverse waves on a string: 
\be
\frac{1}{v^2}\frac{\partial^{2}y}{\partial t^2}=\frac{\partial^{2}y}{\partial x^2},
\ee
is a second-order partial differential equation in \textbf{\emph{both}} time and position.  The solution of this equation is a \textbf{\emph{real}} wave function.  On the other hand, the time-varying wave function satisfying the Schr{\"o}dinger equation is complex.  There is no mystery since, for a first-order linear differential equation of the form
\be\label{1st_order}
\frac{d\psi}{dt}=\alpha\psi,
\ee its solution $\psi(0)e^{\alpha t}$ is non-oscillating \textbf{\emph{unless}} $\alpha$ is an imaginary number.  When the potential $V$ in the Schr{\"o}dinger equation is time invariant, Eq.~\eqref{SE} reduces exactly into two equations one in position and the other in time, of which the latter is in the form of Eq.~\eqref{1st_order}.  The converse is also true, namely, if we insist a wave function to be a complex exponential, then the first-order linear differential equation that gives this wave function as the solution \textbf{\emph{must}} also be complex.  Thus Schr{\"o}dinger had no choice if he chose his differential equation to be first order.

Schr{\"o}dinger has been struggling a lot between choosing a real and a complex wave function.  On the one hand, he expressed his worry to Lorentz about using $i$ by saying that ``What is unpleasant here, and indeed directly to be objected to, is the use of complex numbers.  $\psi$ is surely fundamentally a real function~$\dotsc$\cite{sch_letter_L}''  Five days later he became less reluctant to complex numbers, telling Planck that ``The time dependence must be given by $\psi\sim P.R.\,\left(e^{\pm\frac{2\pi iEt}{h}}\right)$,\cite{sch_letter_P}'' (note however that he takes only the positive real part of the complex exponential) but wrote immediately after that: ``or, what is the \textbf{\emph{same}} [italics mine] thing, we must have $\frac{\partial^{2}\psi}{\partial t^2}=-\frac{4\pi^{2}E^2}{h^2}\psi$,'' which shows his desire of eliminating $i$ (by differentiating $\psi$ twice).   The final Part IV (arrived at Ann. Phys. ten days after he wrote to Planck)\cite{sch_26} of his famous paper inaugurating wave mechanics is suffused with this complex feeling of Schr{\"o}dinger on adopting complex wave functions.  He wrote:
          
\begin{quote}
\textbf{\emph{We will require the complex wave function $\psi$ to satisfy one of these two equations}} [italics Schr{\"o}dinger's].  Since the conjugate complex function $\bar{\psi}$ will then satisfy the \textbf{\emph{other}} equation, we may take the real part of $\psi$ as the \textbf{\emph{real wave function}} [italics mine] (if we require it).\cite{sch_translation}
\end{quote}

and again, as his final conclusion:

\begin{quote}
Meantime, there is no doubt a certain crudeness in the use of a \textbf{\emph{complex}} wave function.  If it were unavoidable \textbf{\emph{in principle}}, and not merely a facilitation of the calculation, this would mean that there are in principle \textbf{\emph{two}} wave functions, which must be used \textbf{\emph{together}} in order to obtain information on the state of the system.  This somewhat unacceptable inference admits, I believe, of the very much more congenial interpretation that the state of the system is given by a real function and its time-derivative.  Our inability to give more accurate information about this is intimately connected with the fact that, in the pair of equations (4''), we have before us only the \textbf{\emph{substitute}}---extraodinarily convenient for the calculation, to be sure---for a real wave equation of probably the fourth order, which, however, I have not succeeded in forming for the non-concervative case.\cite{sch_translation}
\end{quote}

It is not obvious why Schr{\"o}dinger made so much effort to bring complex wave functions into his model despite his seemingly strong reservation.  \textbf{\emph{Perhaps}} the wide adoption of complex exponentials to represent waves by his contemporary colleagues, like Heisenberg,\cite{heisen25} Born,\cite{born&jordan25} and even de Broglie,\cite{deBro25} did generate some ``pressure''.

\section{Conclusion}
Let us conclude by considering the following question.  If taken as an isolated mathematical query, ``what is complex exponential'' has a simple answer.  Ironically, this plain mathematical concept could become overly complex from a physicist's perspective.  To clear the mist we can just ask what is the difference between a real cosine wave, i.e., $\cos(\omega t)$, and the complex exponential $e^{i\omega t}=\cos(\omega t)+i\sin(\omega t)$.  Without effort we realize immediately that adding $ i\sin(\omega t)$ to $\cos(\omega t)$ gives \textbf{\emph{no}} new information.  In the eyes of an electrical engineer, $e^{i\omega t}$ is solely a unit-length vector rotating on a two-dimensional plane with angular velocity $\omega$ as time elapses.  This so-called \textbf{phasor} representation is useful only when he or she is dealing with more than one phasors.  In that case the algebra of complex numbers will simplify much of his or her circuit analysis.  By the same token engineers freely apply any one of the three equivalent forms of Fourier series to fit their problems.  The use of complex exponentials in quantum mechanics does no more than enabling a tidy construct of theory, together with what Schr{\"o}dinger would prefer, ``merely a facilitation of the calculation''.          

\begin{acknowledgments}
I wish to thank my friend and colleague Dr. Cheung Leungfu for his patience in listening my nonsense and clarification of meanings of some German expressions. 
\end{acknowledgments}

\end{document}